\documentclass[%
 reprint,
 superscriptaddress,
 showpacs,
  nofootinbib,
 amsmath,amssymb,
 aps,
 prc,
]{revtex4-1}

\usepackage{verbatim}
\usepackage{graphicx}
\usepackage{epsfig}
\usepackage{amsfonts}
\usepackage{amsmath,amssymb}
\usepackage{bm}
\usepackage{hyperref}
\usepackage{color}

\begin{document}

\title{\bf Shadow poles in coupled-channel problems calculated with Berggren basis}

\author{R. M. Id Betan}
\affiliation{Physics Institute of Rosario (CONICET), 
             Bv. 27 de Febrero 210 bis, S2000EZP Rosario, Argentina}
\affiliation{Department of Physics and Chemistry FCEIA(UNR),
             Av. Pellegrini 250, S2000BTP Rosario, Argentina}

\author{A.T. Kruppa}
\affiliation{Institute for Nuclear Research Hungarian Academy of Sciences (ATOMKI),
             PO Box 51, H--4001, Debrecen, Hungary}

\author{T. Vertse}
\affiliation{Institute for Nuclear Research Hungarian Academy of Sciences (ATOMKI),
             PO Box 51, H--4001, Debrecen, Hungary}
\affiliation{ University of Debrecen, Faculty of Informatics, 
              PO Box 12, H--4010, Debrecen, Hungary}


\begin{abstract}
  \begin{description} 
    \item[Background] In coupled-channel models the poles of the scattering S-matrix are located on different Riemann sheets. Physical observables are affected mainly by poles closest to the physical region but sometimes 
    shadow poles have considerable effect, too.
    \item[Purpose] The purpose of this paper is to show that in coupled-channel problems all poles of the S-matrix can be located by an expansion in terms of a properly constructed complex-energy basis.
    \item[Method] The Berggren basis is used for expanding the coupled-channel solutions.
    \item[Results] The locations of the poles of the S-matrix for the Cox potential, constructed for coupled channel problems, were numerically calculated and compared with the exact ones. In a nuclear physics application the $J^\pi=3/2^+$ resonant poles of $^5$He were calculated in a phenomenological two channel model. The properties of  both the normal and shadow resonances agree with previous findings.
    \item[Conclusions] We have shown that, with an appropriately chosen Berggren basis, all poles of the S-matrix 
     including the shadow poles can be determined. We have found that the shadow pole of $^5$He migrates between Riemann sheets if the coupling strength is varied.
  \end{description}
\end{abstract}

\maketitle

\section{Introduction}
The physics of coupled-channel (CC) models  
spans many research areas, ranging from traditional nuclear physics \cite{atk1,atk2,tho09}
to atomic physics \cite{mar04,nyg06}. It may be extended to chiral perturbation theory combined 
with a multichannel approach \cite{cie13,dot13-a,dot13-b}, and to
hypernuclear physics \cite{miy99}. 

As in  scattering theory in general, in scattering CC models the exploitation of the analytic properties of the S-matrix is a basic tool in describing the scattering processes \cite{ede66}. The S-matrix as a function of energy is analytic over a Riemann surface of many sheets \cite{bad82,gra00,gra03}. The analytic continuation of the physical S-matrix  may have poles on the unphysical sheets.
These poles  correspond to the eigenvalues of the Hamiltonian with various special boundary conditions.
The eigenfunctions belonging to these generalized eigenvalues are referred as resonance/virtual states of the system.
On the physical sheet the eigenfunctions are asymptotically decreasing, i.e., they are bound states. The asymptotic forms of the resonance/virtual states are not bounded. In the following the term resonant state will be used for all discrete 
(bound, resonance and virtual) states.

The main question in a CC scattering model is how the S-matrix poles affect the scattering observables. 
Obviously, they are affected by poles closest to the physical region.
However, it was recognized long ago that poles far from the physical regions (so called 
shadow poles) may have large effects on observables \cite{ede64,fra64}. 
This phenomenon occurs 
in atomic \cite{dij08,rak06,pot88-a,pot88-b}, nuclear \cite{hal87-a,hal87-b,csot93,pea89} and particle physics \cite{mor87-a,mor87-b,mor87-c} as well.
The effect of the shadow poles was also studied for two-dimensional confined electron gas \cite{cat07}.

As for the finding of all types of poles of the S-matrix i.e. solving the CC equation
with a generalized asymptotic condition there are two approaches: either the expansion is modified 
so as to allow for the special boundary conditions, and the problem is solved with orthodox methods
or the problem is transformed so that the usual expansion may provide the modified solutions.. The most straightforward example for 
the second approach is the complex scaling method, which has a long history \cite{cs-a,cs-b}. 
In a recent nuclear three-body application \cite{atk} the back-rotation problem of this  
method has also been solved. 
The usage of the complex scaling method to CC problems sometimes
requires slight extensions, e.g. when shadow poles are sought \cite{csot93a}.

An approach of the first kind nowadays often used is the series expansion of 
the wave function on an extended basis. The basis includes the bound states, 
selected resonance/virtual states as well as complex energy scattering states of the potential \cite{2014IdBetan}.
The existence of such a basis was proven by T. Berggren \cite{Be68}.
Berggren bases were used in the solution of many-body shell model calculations \cite{2002IdBetan,2002Michel}, especially
when weakly bound or unbound  states of certain nuclei observed in radioactive beam facilities
were to be described  by Gamow shell model or complex-energy shell model \cite{Mi09}. 
In certain applications \cite{Bet04,ve89} it was shown that anti-bound 
states can also be included as basis states.

The Berggren bases were also used in CC problems: the radial coupled HFB equations  \cite{mic08} and
the CC Lane equation describing isobaric analog states \cite{Bet08} were solved with it.

The aim of this paper is to show that 
in CC problems {\it all poles} of the S-matrix can be located with properly constructed 
Berggren bases. To this end, we reproduce the results of an exactly solvable CC problem: the one with the Cox potential.
This CC problem can be
solved exactly \cite{1964Cox,pup08,spa06,sam07} so there is no ambiguity in the location 
of the S-matrix poles.
The analytic solution of the Cox potential offers a unique 
opportunity for testing the results of the numerical solution of  
the CC problem.

Another CC problem we consider is the description of the $\frac{3}{2}^+$ resonance state of the nucleus $^5{\rm He}$. This resonance is famous due to its role in the production of thermonuclear energy and in primordial stellar nucleosynthesis. A phenomenological analysis \cite{hal87-a} revealed the presence of a shadow pole. The strong $t+d\to\alpha+n$ transition is accounted for the presence of this pole.
The structure of the  $\frac{3}{2}^+$ state is investigated for example in the works \cite{bro14,ha114,nav11,hal87-b}. Here we apply a simple phenomenological CC model in order to show a physical situation where a shadow pole plays an important role and can be explored using the Berggren basis technique.

The paper is organized as follows. Section~\ref{sec:cox}  
describes the CC Cox problem and its analytic solution. 
In Section ~\ref{sec:berg} we discuss how to use the Berggren basis for the 
solution of a CC problem. Here we discuss the main point of the paper: 
how to choose the Berggren basis in order to get all poles of the S-matrix. Sections \ref{sec:applications} and \ref{sec.he} are dedicated to the applications. The first one presents the results for the CC Cox problem, both in the analytic framework and in numerical frameworks, while the second one contains our study of the pole structure of the $J^\pi=3/2^+$ state of  $^5{\rm He}$ in a two-channel approximation.  Finally, the conclusions  are contained in Section.~\ref{sec:conclusions}.

\section{The Cox coupled-channel problem} \label{sec:cox}
\subsection{The Cox potential}
The two-channel radial Schr\"odinger-equation with energy $E$ in reduced units reads
\begin{equation} \label{eq.h}
 H \psi(r,E) = K^2 \psi(r,E),
\end{equation}
where $K={\rm diag}(k_1,k_2)$, $k_i=\sqrt{E-\Delta_i}$ denotes the 
channel wave numbers, $\Delta_i$ are the threshold energies. We will use
as in \cite{pup08} $\Delta_1=0$ and $\Delta_2=\Delta>0$.
The notation ${\rm diag}(a_1,a_2)$ means a two by two diagonal matrix with elements
$a_i$ in the main diagonal. The Hamiltonian is

\begin{equation}\label{hami}
 H = \left(
  \begin{array}{cc}
    -\frac{d^2}{dr^2} + v_{11}(r) & v_{12}(r) \\
    v_{21}(r) & -\frac{d^2}{dr^2} + v_{22}(r)
  \end{array}
  \right)
\end{equation}
and the solution forms a vector
\begin{equation}\label{solvector}
 \psi = \left(
  \begin{array}{cc}
    \psi_1 \\
    \psi_2 \\
  \end{array}
  \right)~.
\end{equation}
The Schr\"odinger equation in Eq. \eqref{eq.h} has two matrix value Jost solution from which  
the Jost matrix can be constructed defining both the scattering and bound state solutions.
We classify the solutions of Eq. (\ref{eq.h})  as it 
is done in Ref. \cite{pup08}. We call a solution  
a \textit{bound state} when the zero of the determinant of the 
Jost matrix $k_1$ and $k_2$ are both pure positive 
imaginary numbers. We call the solution a \textit{virtual state} 
or \textit{anti-bound state}, when the zero of the Jost-matrix 
determinant corresponds to a real energy below the thresholds and  
the zero is lying on the imaginary 
$k_i$ axes, but not all of them are located on the positive imaginary axis. 
Finally, we call the solution a \textit{resonance} if 
the zero is not lying on any of the imaginary $k_i$ axes, 
hence the corresponding energy is complex or if real 
then it is above at least one of the thresholds.

The derivation of the Cox potential and how to solve it exactly 
are given in \cite{pup08,spa06,sam07}.
To make the paper self-contained we collect some formulas.
The Cox \cite{1964Cox} interaction matrix 
\begin{equation}\label{cox_v}
 V(r) = \left(
  \begin{array}{cc}
    v_{11}(r) & v_{12}(r) \\
    v_{21(r)} & v_{22}(r) \\
  \end{array}
  \right)~.
\end{equation}
is given by 
\begin{equation}
V(r)=-{\cal K}+2 {\cal K}^{1/2}(I+X(r))^{-1}
{\cal K}^{1/2},
\end{equation}
where $I$ is the 2x2 unit matrix and
\begin{eqnarray}
&X(r)={\rm diag}(\exp(-\kappa_1r),\exp(-\kappa_2 r)) X_0\nonumber\\
&\times {\rm diag} (\exp(-\kappa_1 r),\exp(-\kappa_2 r)).
\end{eqnarray}

The symmetric 2x2 
matrix $X_0$ contains the parameters of the potential. The factorization wave numbers $\kappa_1$ and $\kappa_2$ are positive parameters and they
satisfy the condition $\kappa_2^2-\kappa_1^2=\Delta.$  (we will use $\kappa_1=\kappa$ as
independent parameter). The matrix $\cal K$ is related to the factorization energies by 
${\cal K}={\rm diag}(\kappa_1,\kappa_2)$. Of course, the interaction matrix
\eqref{cox_v} is symmetric.

The determinant of the Jost matrix is given by \cite{pup08}
\begin{equation} \label{eq.f}
 f(k,p)=\frac{(k+i\, \alpha_1)(p+i\, \alpha_2)+\beta^2}{(k+i\, \kappa_1)(p+i\, \kappa_2)}.
\end{equation}
Here we denote for convenience the channel wave numbers as $k_1=k$ and $k_2=p$.
The threshold condition reads $k^2-p^2=\Delta$. 
The zeros of the function $f(k,p)$ determine the position of 
the bound, virtual and resonance states.
Interestingly the Cox potential depends on the factorization wave numbers but the 
eigenenergies are independent from $\kappa_i$.

The connection between the
parametrization $\alpha_1$, $\alpha_2$, $\beta$  and $X_0$ is given by the equation
\begin{equation}
  U_0=\left(\begin{array}{cc}
    \alpha_1 & \beta \\
    \beta & \alpha_2 \\
  \end{array}\right)={\cal K}^{1/2}(I-X_0)(I+X_0)^{-1}{\cal K}^{1/2}
\end{equation}
and the inverse relation is 
\begin{equation}
X_0={\cal K}^{-1/2}({\cal K}-U_0)({\cal K}+U_0)^{-1}{\cal K}^{1/2}.
\end{equation} 

\subsection{Reduced inverse problem\label{sec.inverse}}
In a \textit{direct problem} we calculate the eigenenergies and the
corresponding channel wave numbers 
$(k_i,p_i)$ of the Cox 
CC equations for a given set of potential parameters 
 $\alpha_1$, $\alpha_2$, $\beta$, and $\Delta$.
In the \textit{inverse problem} we fix a few eigenenergies or other
characteristics of
the problem and search for the parameters of the potential 
which give back the fixed characteristics. 

A very convenient 
approach was  introduced in Ref. \cite{2007Pupasov}. 
In this  approach one first fixes two solutions and 
the other two solutions are obtained in closed form.
Let us fix $(k_1,p_1)$ and $(k_2,p_2)$ as 
the first two zeros of the Jost-matrix determinant $f(k,p)$
then from \eqref{eq.f} we get 
\begin{eqnarray}
 (k_1+i\, \alpha_1)(p_1+i\, \alpha_2)+\beta^2 &=&0 \\
 (k_2+i\, \alpha_1)(p_2+i\, \alpha_2)+\beta^2 &=&0
\end{eqnarray}
and we obtain the parameters 
$\alpha_1$ and $\alpha_2$ in terms of $(k_1,p_1)$ and $(k_2,p_2)$ and $\beta$

\begin{eqnarray}\label{alpha_sol1}
  \alpha_1 &=& \frac{1}{2} \left[ i(k_1+k_2) \pm \sqrt{-\Delta^2_k - 4 \beta^2 \frac{\Delta_k}{\Delta_p}} \right] \label{eq.a1} \\
  \alpha_2 &=& \frac{1}{2} \left[ i(p_1+p_2) \mp \sqrt{-\Delta^2_p - 4 
  \beta^2 \frac{\Delta_p}{\Delta_k}} \right] \label{eq.a2},
\end{eqnarray}
where $\Delta_k = k_2 - k_1$ and 
$\Delta_p = p_2 - p_1$.
The signs define 
two sets of solutions which correspond to the same $(k_1,p_1)$ and $(k_2,p_2)$ but 
two different $(k_3,p_3)$ and $(k_4,p_4)$ roots. In selecting  
the solution either the upper or the lower signs are to be taken.

The other two roots $(k_3,p_3)$ and $(k_4,p_4)$ are determined in Ref. \cite{2007Pupasov}
as follows.
\begin{eqnarray} 
  k_3 &=& \frac{1}{2} \left[ \mp i \sqrt{-\Delta^2_k - 4 \beta^2 \frac{\Delta_k}{\Delta_p}} + \sqrt{D_k} \right] \label{eq.k3} \\
  p_3 &=& \frac{1}{2} \left[ \mp i \sqrt{-\Delta^2_p - 4 \beta^2 \frac{\Delta_p}{\Delta_k}} + \sqrt{D_p} \right] \label{eq.p3} \\
  k_4 &=& \frac{1}{2} \left[ \mp i \sqrt{-\Delta^2_k - 4 \beta^2 \frac{\Delta_k}{\Delta_p}} - \sqrt{D_k} \right] \label{eq.k4} \\
  p_4 &=& \frac{1}{2} \left[ \mp i \sqrt{-\Delta^2_p - 4 \beta^2 \frac{\Delta_p}
  {\Delta_k}} - \sqrt{D_p}\right] \label{eq.p4},
\end{eqnarray}
where  
  $D_k = \Delta^2_k + 4 \beta^2 \frac{\Delta_p}{\Delta_k} + 4 k_1 k_2$ and
$D_p = \Delta^2_p + 4 \beta^2 \frac{\Delta_k}{\Delta_p} + 4 p_1 p_2$.
Note the difference in sign of the second term in $p_3$ and $p_4$ with respect to Eq. (50) in Ref. \cite{2007Pupasov}.

\section{Solution using Berggren basis}\label{sec:berg}
We calculate the eigenvalues of the CC Cox problem by
diagonalizing the Hamiltonian (\ref{hami}) in the Berggren 
bases of the potentials $v_{11}(r)$  and $v_{22}(r)$. 
We consider two auxiliary problems
\begin{equation}\label{ham_bas}
\left [-\frac{d^2}{dr^2}+v_{ii}(r)-E_n^{(i)}\right ] u^{(i)}_n(r)=0\ \ \ i=1,2.
\end{equation}

The states of the Berggren basis are the solutions of Eq. \eqref{ham_bas} 
and  
for each channel they are composed of the resonant basis states which 
are eigensolutions of Eq. (\ref{ham_bas}) with 
purely outgoing wave boundary condition, i.e. they correspond to the poles 
of the $S$-matrix in that channel.
The resonant states (bound,virtual and resonance solutions) with energy $E_n^{(i)}$ are 
denoted by $u^{(i)}_n(r)$. 
Beside the resonant  
states the basis contains scattering states along a 
complex contour $L$.
The scattering solutions are denoted by
$u^{(i)}(r,E)$ or $u^{(i)}(r,k)$ if we use the energy $E$ or
wave number $k$, respectively.

The shape of the contour $L$  is restricted by certain rules \cite{Be68}. In the $k$ plane the contour
has to go through the origin
and has to be symmetric to the origin, i.e. if $k$ is on the contour $L$ 
then $-k$ should be on the contour $L$ too.
Contour $L$ is divided by the origin to two halfs, denoted by $L^+$ and $L^-$.
The half  $L^+$ is between $k=0$ and $k\rightarrow+\infty$ with $Im(k)<0$ in Figure 1 and Figure 2. The  $L^+$ part  for large $|k|$ values  
 has to go back to the real axis and remain there. The contribution of the
 $L^-$ half is equal to the one of $L^+$ and the factor $2$ is included in the normalization of the scattering wave functions.
When the contour is the full real $k$ line similar decomposition of the $k$ values is used in \cite{newton} proving the completeness of the scattering states.

\begin{figure} 
  \includegraphics[width=1.\columnwidth]{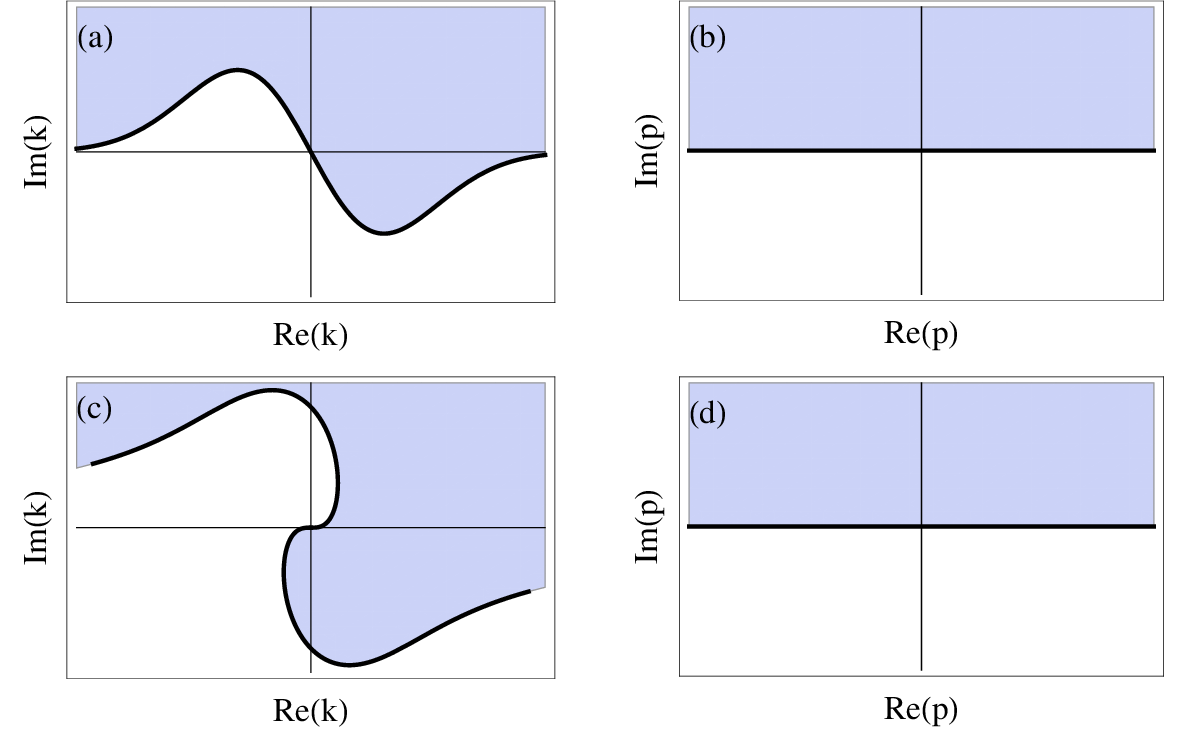}
  \caption[T]{\label{r2sheet}
  (Color online) Illustration of the contours $L$ (thick lines) on the $k$ and $p$ planes 
  for the second Riemann sheet.
  Resonance states  can be determined with contours similar to the
  parts (a) and (b). Both virtual and resonance states can be calculated with contours 
  similar to the parts (c) and (d). The double roles of the shaded areas are explained in the text.}
\end{figure}

\begin{figure} 
  \includegraphics[width=1.\columnwidth]{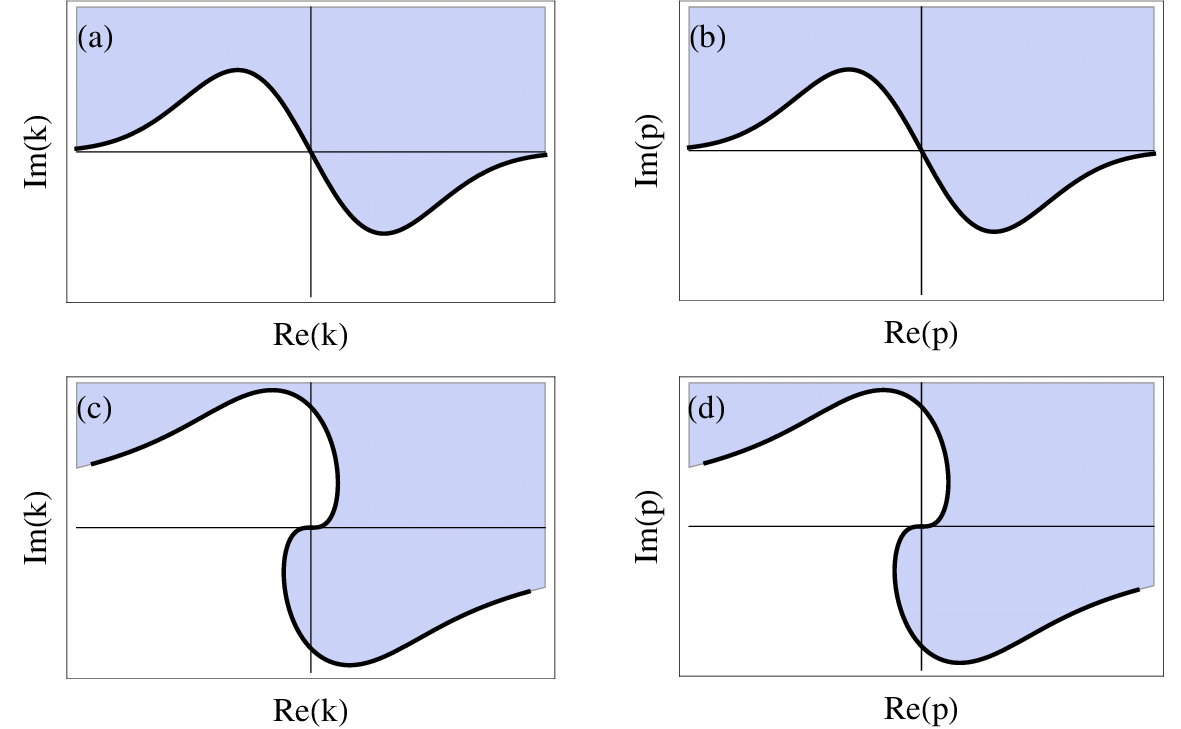}
  \caption[T]{\label{r3sheet}
  (Color online) Similar to Fig. \ref{r2sheet} but for the third Riemann sheet.}
\end{figure}

In the set up of the basis only those resonant states 
have to be included into the Berggren basis whose wave numbers are in the shaded area, i.e. between the contour $L^+$ and the positive $k$-axis. Similar relations should be hold for the $p$ 
contour of the second channel. (See Figures \ref{r2sheet} and \ref{r3sheet} the wave numbers 
to be included into the basis
should be also in the
shaded area of the figures.)

The completeness relation of the Berggren basis reads 
\begin{eqnarray}\label{eq:delb}
 \delta(r-r^\prime)=\sum_{n=b,d,v}{u^{(i)}_n(r)}u^{(i)}_n(r^\prime) \nonumber \\
 +\int_{L^+} dk~ {u^{(i)}(r,k)}u^{(i)}(r^\prime,k)~.
\end{eqnarray}
In this relation (and later) the notation $n=b,d,v$ means that the sum over 
$n$ runs through all  bound states,
decaying resonances and virtual states in the shaded area of Fig. \ref{r2sheet}.
The integral in 
Eq. (\ref{eq:delb}) is over the
scattering states along the  $L^+$ half contour. 
The completeness relation in Eq. (\ref{eq:delb}) for chargeless particles was introduced 
in Ref. \cite{Be68}
and its validity  has been shown for charged particles too \cite{Michel08,Michel,akram}

Since we can not handle the continuum part exactly  the complex contour is 
discretized and truncated in order to have a finite number of contour 
states. The renormalization of the discretized contour states was introduced
first in Ref. \cite{Li96} and
it is performed as  in Ref. \cite{Mi09}. 
We use as discretization points $E_k^{(i)}$ ($k=1,\ldots,N_i$) the abscissas of a Gaussian
quadrature procedure. The corresponding weights of the quadrature points 
are denoted by $h^{(i)}_k$. After discretizing the integral in Eq. (\ref{eq:delb}) 
an approximate completeness relation for the finite number of basis
states reads
\begin{equation}\label{eq:finb}
 \delta(r-r^\prime)\approx \sum_{n=b,d,v,c}^{M_i} {w^{(i)}_n(r,E^{(i)}_n)}w^{(i)}_n(r^\prime,E^{(i)}_n),
\end{equation}
where $c$ labels the discretized scattering states from the contour $L$ and $M_i$ is 
the sum of the resonant (bound, virtual and resonant) states contained inside the 
the shaded area 
plus $N_i$ number of discretized continuum states. 
If $E^{(i)}_n$ is a scattering energy from the contour $L$ then 
the scattering state of the discretized continuum is denoted by 
$w^{(i)}_n(r,E^{(i)}_n)=\sqrt{h^{(i)}_n} u^{(i)}_n(r,E^{(i)}_n)$. If however $E^{(i)}_n$ 
corresponds to a normalized resonant state of the potential $v_{ii}(r)$ then 
$w^{(i)}_n(r,E^{(i)}_n)= u^{(i)}_n(r)$. 
The set of Berggren vectors form a bi-orthonormal basis in the truncated space
\begin{equation}\label{eq:biort}
 <\tilde {w}^{(i)}_n|w^{(i)}_m>=\delta_{n,m}~.
\end{equation}
with $<\tilde{w}^{(i)}_n|r>=<r|w^{(i)}_n>=w^{(i)}_n(r,E^{(i)}_n)$.

Having fixed the Berggren basis the solution \eqref{solvector}
is approximated in the form
\begin{equation}
\psi_i(r)=\sum_{k=1}^{M_i}C^{(i)}_kw^{(i)}_k({
r},E^{(i)}_k)\ \ \  i=1,2.
\end{equation}
Using Eq. (\ref{eq.h}) we get the following set of linear equations 
for $C^{(i)}_k$ 

\begin{eqnarray}\label{eq:secul1}
&(E^{(1)}_k-E)C^{(1)}_k+\sum_{m=1}^{M_1}\langle\tilde w^{(1)}_k|
v_{12}|w^{(2)}_m\rangle C^{(2)}_m=0\nonumber \\
&k=1,\ldots,M_1
\end{eqnarray}

and

\begin{eqnarray}\label{eq:secul2}
&(E^{(2)}_k-(E-\Delta))C^{(2)}_k+\sum_{m=1}^{M_2}\langle\tilde w^{(2)}_k|
v_{21}|w^{(1)}_m\rangle C^{(1)}_m=0\nonumber\\ 
&k=1,\ldots,M_2.
\end{eqnarray}

These two equations can be combined into one matrix eigenvalue equation. 
By diagonalizing the matrix of the Hamiltonian we get complex 
eigenvalues $E_\nu$ $\nu=1,\ldots M_1+M_2$.
Some complex/real eigenvalues $E_\nu$ can be identified as resonant states of the
CC problem. The identification  in this case is easy because
we should find the $E_\nu$ eigenvalue being closest to the exact value.
In general this task is more complicated, some methods can be found in
Ref.\cite{Mi09} . 

In two-channel case we have a Riemann surface with four sheets. Let us define the  
four Riemann sheets in terms of the sign of the imaginary parts of the $k$ and the $p$ 
wave numbers.
The Riemann sheets can be labeled by a two-term sign string $({\rm sgn}({\rm Im} k),{\rm sgn}({\rm Im} p))$. 
We follow the standard notations introduced in Refs. \cite{fra64,bad82}. The first sheet is 
the physical  one and it is signed by $(+,+)$. The second sheet is $(-,+)$
and these two levels are connected if $0<E<\Delta$. The third and fourth sheets are identified by 
$(-,-)$ and $(+,-)$, respectively. These two sheets are also connected if $0<E<\Delta$.  If the energy $E$ is
above the  threshold $\Delta$ the topological structure changes: sheets one and three as well as  
two and four are connected \cite{fra64}. The location of a resonant state determines the
asymptotic behavior of its wave function. 
Bound state from the first Riemann sheet has square integrable wave functions
in both channels. However a resonant state from the second Riemann sheet have such a wave
functions that the first component asymptotically diverges and the second channel 
has bound state type 
behavior. For resonant states from
the third Riemann level both components of the wave function diverge asymptotically.

When we diagonalize  the CC Cox potential  in  
Berggren bases sometimes we have to take different contours in 
the complex $k$ and $p$ planes in order to determine 
the solutions we are interested in. 
As we discussed earlier the shape of the complex  contour $L$ determines which resonant  
states of the potential $v_{ii}(r)$ should be included into the Berggren basis of the given
channel.
However the shape of the contours also 
determines the Riemann sheets and we are able to find only 
the resonant CC states on that Riemann sheets. The CC resonant states of a given calculation 
are in the shaded areas both in the $k$ and $p$ planes. 

If both the $k$ and the $p$ contours remain on the real axis we can find 
only bound states on the physical $(+,+)$ sheet. In order to locate resonant states 
on the second 
Riemann sheet $(-,+)$  we have to use contours of the form displayed on Fig. \ref{r2sheet}.
Contours similar to the upper part ((a) and (b))can be used only for calculation of 
resonance states of the second Riemann
sheet and for bound states of  the first Riemann sheet. If a virtual 
state is located on the second Riemann level then the contours have to look like as 
displayed on the lower
part ((c) and (d)) of Fig. \ref{r2sheet}. Of course, also resonance states on the second Riemann sheet 
can be calculated using 
contours similar to the lower part ((c) and (d)) of Fig. \ref{r2sheet}. This type of contours 
however discard some CC bound state from the first Riemann level.

If resonant states located on the third
Riemann sheet are to be determined, the shape of the contours  
depicted in Fig. \ref{r3sheet} have to be used. 
Only resonance
states can be determined by contours similar to the upper part ((a) and (b) of Fig. \ref{r3sheet}.
The lower part is appropriate for calculation aimed at obtaining virtual states and resonance 
states located on the third Riemann level. We mention that using contours 
corresponding to the upper part ((a) and (b)) of Fig. \ref{r3sheet}  bound and resonance states of all 
Riemann sheets 
can be determined simultaneously. However numerically it is favorable to use simpler contours.
For resonant state on the second Riemann sheet the accuracy of the numerical calculation is 
better for contours on Fig. \ref{r2sheet} than for contours of Fig. \ref{r3sheet}.
Simpler contours for states located on the fourth Riemann sheet can be similarly
constructed.

\section{Application: Cox potential} \label{sec:applications}
At certain parameters of the Cox potential the two channels decouple.
Since the eigenvalue problem is exactly solvable, the accuracy of our program for
solving the single-channel problem can be conveniently checked.
This program integrates the differential equation
in Eq.(\ref{ham_bas}) numerically. It is important that we solve the single-channel problem accurately, since the discrete basis states are calculated using this program. The inaccuracy of the basis states would spoil
the numerical results of the CC system. Therefore 
we deal with the solution of the single-channel case first. Then we will consider a CC system having a resonance state
above the first threshold and below the second one. 
The threshold $\Delta$ and $\beta$ are fixed to the values $1$ and $0.1$, 
respectively for the applications considered here.

\subsection{Single-channel solutions}
If we take the parameter $\beta=0$ then the CC equation with 
Cox potential reduces to two  uncoupled equations. In this case the zeros of 
the Jost determinant (\ref{eq.f}) are $k=-i\, \alpha_1$ and $p=-i\, \alpha_2$. 
We compare the analytical results with that of our numerical 
calculations in Table \ref{table.uncop}. 
Note that the exact value of ${\rm Im}(k)$ has opposite sign 
than the value of the parameter $\alpha_1$.

\begin{table}[ht] 
\begin{ruledtabular}
\caption{\label{table.uncop} Wave numbers of the anti-bound/bound states calculated exactly 
and numerically for the single-channel problem with potential $v_{11}(r)$ using $\beta=0$ and $\kappa=1$.}
  \begin{tabular}{ccc}
 $\alpha_1$ & ${\rm Im}(k)$ exact & ${\rm Im}(k)$ numerical \\
\hline
 0.7      & -0.7          & -0.699528     \\
 0.6      & -0.6          & -0.598954     \\
 0.5      & -0.5          & -0.499935     \\
 0.4      & -0.4          & -0.399984     \\
 0.3      & -0.3          & -0.299996     \\
 0.2      & -0.2          & -0.199998     \\
 0.1      & -0.1          & -0.099998     \\
-0.1      & +0.1          &  0.100001  \\
-0.2      & +0.2          &  0.200000  \\
-0.3      & +0.3          &  0.299999  \\
-0.4      & +0.4          &  0.399999
  \end{tabular}
  \end{ruledtabular}
 \end{table}
To calculate the eigenvalues we used the highly reliable Fortran program ANTI \cite{Ix95} 
which is based on Ixaru's method \cite{Ix84} for the numerical 
solution of the differential equation \eqref{ham_bas}. 
This program reproduces the exact results reasonably well in most of the cases given 
in Table \ref{table.uncop}.
The agreements are best for the bound state cases and the anti-bound wave numbers are also 
reproduced well, although the deviation from the exact value has increased gradually 
as the $\alpha_1$ value has increased.
In solving numerically the problem the diagonal potentials 
$v_{11}(r)$ and $v_{22}(r)$ are cut to zero at a reasonable large  
$R_{{\rm max}}$. 
Beyond $R_{{\rm max}}$ the potential is considered to be zero.
The results should be at most slightly dependent on the chosen value of $R_{{\rm max}}$. 
With a cut-off radius  $R_{{\rm max}}=13$ our numerical result is  
${\rm Im}(k)=-0.699528$ for $\alpha_1=0.7$, which 
deviates from the exact value in the fourth decimal digit. 
With the same $\alpha_1$ value if we changed the cut-off radius value to 
smaller or larger values we got slightly different ${\rm Im}(k)$ values.
(For $R_{{\rm max}}=12$ we got ${\rm Im}(k)=-0.700028$.  For $R_{{\rm max}}=14$ we got ${\rm Im}(k)=-0.698944$.)
So we found that the wave number of the anti-bound state depends only weakly on the
cut-off radius of the diagonal potential. This is in agreement with the finding
in Ref.\cite{Da12}  for a cut-off Woods-Saxon potential.
The pole energy of the $S$-matrix is determined from the condition that
the logarithmic derivatives of the internal and the external solutions of the
equation(\ref{ham_bas}) are equal at a matching distance $R_{{\rm match}}$.
See e.g. Ref.\cite{Da12,Ix95,Ve82}.
The internal solution is regular in $r=0$, while the external solution
is a purely outgoing wave at $R_{{\rm max}}$. In principle the pole energy
should not depend on $R_{{\rm match}}$. 
In our calculation the value of $R_{{\rm match}}$ influenced only the fifth decimal digit of ${\rm Im}(k)$ 
if we used a value in the range $R_{{\rm match}}\in [1,5]$. 

These comparisons of the exact and approximate 
energies give some hint on the limits of accuracy we can expect 
between the exact and approximate results of the CC 
calculations. 
We certainly can not expect better agreement 
for the CC case than we got for the single-channel case.

\subsection{Coupled-channel: exact solutions}
In order to obtain the exact solution to the CC problem we will appeal to 
the inverse procedure introduced in section \ref{sec.inverse}. Let us consider 
a resonance solution of the Cox potential with complex energy $E_r-iE_i$ so that $0 < E_r < \Delta$ and 
$E_i>0$. We will determine $(k_1,p_1)$ 
and $(k_2,p_2)$ from the complex energy solutions 
$E_1=E_r - i E_i$ and $E_2=E_1^*=E_r + i E_i$, 
which correspond to the wave numbers $k_1=k_r + i k_i$ and $k_2=-k^*_1=-k_r + i k_i$ 
with $k_r>0$ and $k_i<0$. The relations between the real and imaginary parts of 
the energy and wave numbers are 

\begin{eqnarray}
  k_r &=& \frac{1}{\sqrt{2}}   \left[ E_r + \sqrt{E_r^2 + E_i^2} \right]^{1/2} \\
  k_i &=&-\frac{E_i}{\sqrt{2}} \left[ E_r + \sqrt{E_r^2 + E_i^2} \right]^{-1/2} 
\end{eqnarray}
(note that Eq. (52) of Ref. \cite{2007Pupasov} is wrong). 
Using the threshold condition $k^2-p^2=\Delta$ we can determine $p_r$ and 
$p_i$ with $p_1=p_r + i p_i$ and $p_2=-p^*_1=-p_r + i p_i$. The sign of $p_r$ is 
determined by noticing that from Eq. (31) of Ref. \cite{2007Pupasov} 
we can get  $k_r p_r<0$, while the sign of $p_i$ is determined by the 
condition $p_r p_i<0$ for $E_i>0$ (which also implies $k_i p_i<0$). 
Considering these restrictions we have

\begin{eqnarray}
  p_r &=& \frac{-1}{\sqrt{2}}  \left[-(\Delta-E_r) + \sqrt{(\Delta-E_r)^2+E_i^2} \right]^{1/2} \\
  p_i &=& \frac{E_i}{\sqrt{2}} \left[-(\Delta-E_r) + \sqrt{(\Delta-E_r)^2+E_i^2} \right]^{-1/2}.
\end{eqnarray}

Taking for example $E_r=0.4$ and $E_i=0.01$ as in \cite{2007Pupasov} we get
$k_r=0.632550$, $ k_i=-0.007905$ 
and $p_r=-0.006455$ $p_i=0.774624$.
The exact solutions for the upper sign in Eqs. 
(\ref{eq.k3}--\ref{eq.p4}) 
are displayed in Table \ref{table.cor}. Now the CC problem has two resonances and two 
anti-bound states.
According to Table \ref{table.cor} the resonances $E_1$ and $E_2$ 
are located in the second Riemann sheet. The anti-bound $E_3$ state is on the
second sheet too while the second anti-bound solution $E_4$ is a shadow pole on the third Riemann sheet.

\begin{table} 
\begin{ruledtabular}
\caption{\label{table.cor} Exact energies ($E_j$ for $j=1,\dots ,4$) and
corresponding wave numbers in the CC Cox 
potential with 
parameters given in the text.}
  \begin{tabular}{ccc}
$E_j$ & $k_j$ & $p_j$ \\
\hline
0.4-i0.1& 0.632550 -i\, 0.007905&-0.006455 +i\, 0.774624\\
0.4+i0.1&-0.632550 -i\, 0.007905& 0.006455 +i\, 0.774624\\
-0.5604738&-i\, 0.748648&i\, 1.24919\\
-0.5995714&-i\, 0.774302&-i\, 1.26473\\
\end{tabular}
\end{ruledtabular}
\end{table}

\subsection{Coupled-channel: approximate solutions\label{sec:path}}
The approximate solutions are obtained by direct diagonalization of the Cox potential
on Berggren basis as it is described in section \ref{sec:berg}.
For this  we need to find the parameters $\alpha_1$ and $\alpha_2$  of the potential
which give the energies of the exact solution of the previous section. 
Using Eqs. \eqref{eq.a1} and \eqref{eq.a2} and the upper signs we get
$\alpha_1=0.769379934$, $\alpha_2=-0.766852669$.

\begin{table}[ht]
\caption{\label{table.spe} Energies and wave numbers of discrete basis states
in the first and in the second channel.}
\begin{ruledtabular}
\begin{tabular}{ccc}
  channel& Re$E^{(i)}_n$ & Im$k^{(i)}_n$ \\
\hline
  1      & -0.573887     &-0.757553\\
  2      & -0.592006     & 0.769419
\end{tabular}
\end{ruledtabular}
\end{table}

The solutions of CC equations using  
Berggren bases are carried out as follows. Two 
Berggren bases are calculated using the diagonal potentials $v_{11}(r)$ and $v_{22}(r)$, 
respectively, with the  $\kappa=1$ parameter value. 
The parameter $\kappa$ affects the shape of the radial 
potentials but does not affect the CC eigenenergies. 
For $\kappa=1$ we found that the unperturbed potential $v_{11}(r)$ has
an anti-bound state 
and the unperturbed potential $v_{22}(r)$ has a bound state. 
The actual 
values of the energies and wave numbers of the resonant basis states are given in Table \ref{table.spe}.

The decaying CC resonance state at the energy $E_1=0.4\, -\, i\, 0.01$ 
is on the second Riemann sheet
.
The shape of the contour therefore should look like as the upper part ((a) and (b)) of Fig.
\ref{r2sheet}, 
since the contour should go below $k_1$ 
and  $k_1$ should be in
the shaded area.

\begin{figure} 
   \includegraphics[angle=-90,width=0.8\columnwidth]{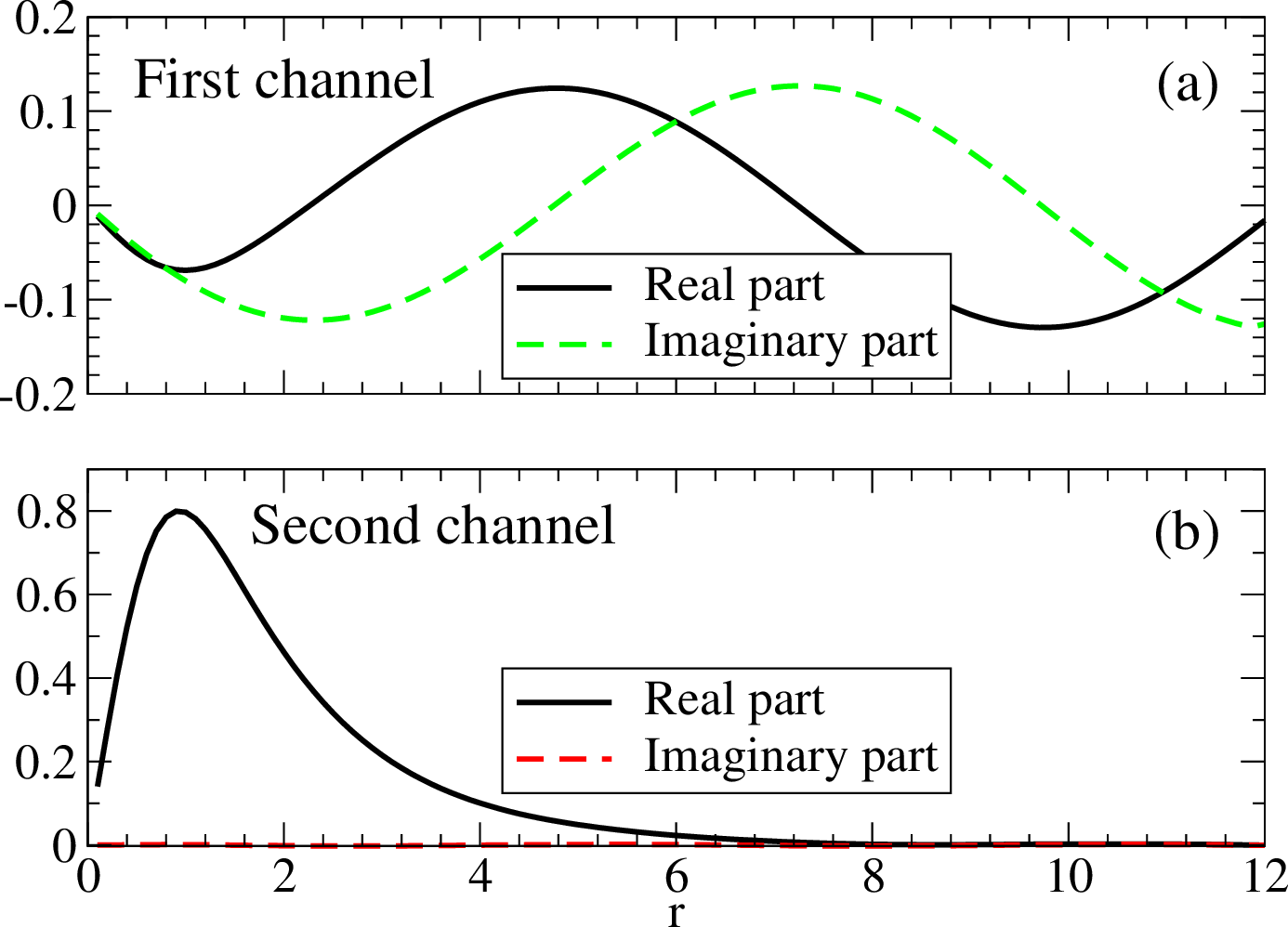}
  \caption[T]{\label{wffig}
  The wave function of the first resonance state located in 
  the second Riemann sheet and calculated with the Berggren expansion method. 
  The upper (a) and lower (b) parts show
  the first and second channel components, respectively. 
  The imaginary part of the second component
  is practically zero.}
\end{figure}

The basis in the first channel has no bound state therefore it is formed 
from the complex continuum 
states only. In the  second channel the basis contains the unperturbed 
bound state and a continuum which can be taken along the real $p$-axis.
Diagonalizing the Cox potential using these bases, 
we got a decaying resonance at the energy $E=0.400047 - i\, 0.0100011$. 
The corresponding wave function is displayed in Fig. \ref{wffig}. The form of the wave
function follows from the rules discussed in section \ref{sec:berg}. 
The real and the imaginary parts of the first channel 
wave function show resonant behavior, i.e. they both diverge asymptotically. 
The real and the imaginary parts 
of the second channel wave function however both falls asymptotically as a bound state wave 
function does.

Beside the resonance states at $E_1$ and $E_2=E_1^*$ there is an anti-bound state at the 
energy $E_3=-0.5604738$. It is also in the second Riemann sheet. In order to 
calculate this state
the contour should be taken similar to the one in the lower part ((c) and (d)) of the 
Fig. \ref{r2sheet}, since $k_3$ should be in the shaded area therefore we had to modify the contour used before.
We still have two possibilities for selecting the basis in this channel. If the contour crosses the imaginary $k$-axis far from the origin, say at $k=(0,-1.2)$
then the anti-bound basis state will be in the shaded area and should be included
into the Berggren basis. Therefore the 
 Berggren basis in the first channel contains the unperturbed anti-bound 
state at $k_n=(0,-0.757553)$
and a set of discretized complex $k$ scattering states and in the 
second channel the unperturbed bound state and the 
real $p$ scattering states are in the basis.
If we use this basis the diagonalization of the Cox potential gives a CC
virtual state at energy 
$-0.561467 - i\, 0.494 \times 10^{-7}$  which is 
very close to the exact $E_3$ value.
The other option for choosing the contour is that we cross the imaginary $k$-axis just between 
the exact $k_3=(0,-0.749648)$ and the unperturbed anti-bound state at $k_n=(0,-0.757553)$. If we 
cross the imaginary axis at $(0,-0.75)$ then the unperturbed anti-bound state will be outside the shaded
area therefore it won't be included in the basis. By using this basis in the
first channel (the basis in the second channel remains unchanged) 
the diagonalization gives a CC
virtual state at energy 
$-0.561008 - i\, 0.413 \times 10^{-3}$  which is also
very close to the exact $E_3$ value.
This later basis shows an example for a case in which a correlated anti-bound state is produced by diagonalization with Berggren bases in which only bound state and complex scattering
states are included. The small imaginary parts of the energies in the results
of the diagonalization in both cases are due to the numerical errors of the
numerical procedures used. They are beyond the accuracies of the errors of the single channel calculation of the anti-bound basis state for $\alpha_1=0.8$.

In the Cox potential there is an another anti-bound solution at the energy
$E_4=-0.599544$.
This state lies on the third
Riemann sheet so it is a shadow pole.
To be able to expand this state we have to use a contour which is similar to the one
in the lower part ((c) and (d)) of Fig. \ref{r3sheet}.
The Berggren basis in the first channel contains the unperturbed  anti-bound state. 
Because of the symmetry requirement of the complex $L$ 
 contour to the origin,  in the second channel
the unperturbed bound state
should be excluded from the basis. Now we have no alternative in
choosing the contour since $k_4=(0,-0.774302)$ lies lower than the negative of the imaginary
part of the bound state at $-k_n^{(2)}=-0.769419$. The bound pole now is in the not shadowed part.  But a finite number of 
discretized scattering states with complex $p$ wave number  
are naturally included in the basis. The diagonalization of the Cox 
potential in these bases gives an anti-bound CC state at the 
energy $-0.600357 - i\, 0.149 \times 10^{-4}$, which is very close to 
the exact $E_4$ value. 
The deviation from the exact value is again within the accuracy of reproducing
the exact single particle basis states.

The Berggren basis allows the calculation of more than 
one resonant states simultaneously with a properly chosen basis. 
As an interesting example we consider the first resonance state from the second Riemann
sheet and the shadow 
anti-bound state from the third Riemann sheet. Numerically we will show that 
although the states are
on different sheets they can be calculated simultaneously by using the same
bases.
Because we want to calculate an anti-bound state in the third Riemann sheet 
we have to use contours
similar to the one in the lower part ((c) and (d)) of Fig. \ref{r3sheet}. As we discussed before these contours 
however might exclude
some resonance states located in the second Riemann sheet. 
Therefore we have to choose  the
crossing points of the contours and the positive imaginary axes carefully, 
since the wave numbers of the 
resonance CC state 
should be in the shaded area of Fig. \ref{r3sheet}.
The Berggren basis in the first channel is formed by 
the unperturbed anti-bound state and a set of   scattering states along the $k$
contour, while  in the second 
channel the basis is composed of the unperturbed bound state and a 
set of complex $p$ scattering states. The 
diagonalization of the Cox potential using these bases gives the following eigenenergies 
$0.400214 - i\, 0.0098868$ and 
$-0.599100 - i\, 0.501 \times 10^{-3}$ simultaneously. 
Although these numerical results received by these bases are still quite close to 
the exact values of $E_1$ and $E_4$,   the quality of the approximation
is a little bit poorer than the ones we presented earlier with bases adjusted
to the energies individually.
Nevertheless the accuracies are still inside the ones of the single channels
basis states.

\section{Application to nuclear physics: a shadow pole in $^5$He} \label{sec.he}
In this section we are going to study the structure of the $J^\pi=3/2^+$ state of the $^5$He nucleus in the two-channel model \cite{1966Facio}. Here the CC Schr\"odinger equations are given by
\begin{eqnarray}
 \left[ h_1 - (E-\Delta_1) \right] \, u_1(r) + V_{12}\, u_2(r) &=& 0, \\
 \left[ h_2 - (E-\Delta_2) \right] \, u_2(r) + V_{12}\, u_1(r) &=& 0, 
\end{eqnarray}
where channel one describes the $^4$He-$n$ partition, and channel two describes the $^3$H-$d$ partition. For thresholds we take $\Delta_1=0$, and $\Delta_2=17.59$ MeV \cite{tunl}.

The two positive-parity channel states coupled through the interaction $V_{12}$ are $^2D_{3/2}$ and $^4S_{3/2}$, with the following single-particle Hamiltonians $h_1$ and $h_2$, respectively 
\begin{eqnarray}
  h_1(r) &=& - \frac{\hbar^2}{2\mu_1} \frac{d^2}{dr^2}
           + \frac{\hbar^2}{2\mu_1} \frac{l_1(l_1+1)}{r^2}  
           + V_1(r) + \bar{l}_1 \cdot \bar{s}\, V_{so}(r) \nonumber  \\
  h_2(r) &=& - \frac{\hbar^2}{2\mu_2} \frac{d^2}{dr^2}
           + V_{coul} +  V_2(r)  + V_{sws}(r) \nonumber
\end{eqnarray}
with $l_1=2$, and $\mu_1=0.805686$ amu and $\mu_2=1.205288$ amu, the reduced masses of the$^4$He-$n$ and $^3$H-$d$ fragmentations, respectively.

The central potential $V_1(r)$ and the spin-orbit term $V_{so}(r)$  for the first channel, unlike in Ref. \cite{1966Facio} contain no repulsive cores,
\begin{eqnarray}
   V_1(r) &=& \frac{-V_1}{1 
           + \left( \frac{r}{a_1}-1 \right) e^{\frac{r-R_1}{a_1}}} \\ 
  V_{so}(r) &=& \frac{a_1^2}{r}\, \frac{V_{so}}{V_1}\, \frac{dV_1(r)}{dr} \nonumber   \\
\end{eqnarray}
with $V_1=70.13$ MeV, $V_{so}=15.0$ MeV, $a_1=0.85$ fm, and $R_1=1.70$ fm. Since we droped the hard core, we had to adjust the strength $V_1$ in order to reproduce the resonance parameters $E_r=0.798$ MeV, and $\Gamma=0.648$ MeV of ground state $3/2^-$ of $^5$He. Using these parameters, the resonant ground state is found at $\varepsilon(p_{3/2})=0.799-i\, 0.361$ MeV. The energy of the state $d_{3/2}$ is found at $\varepsilon(d_{3/2})=15.4-i\, 28.0$ MeV, which is a physically irrelevant resonance since the imaginary part of the energy  is bigger than its real part.
Since the Hamiltonian $h_1(r)$ does not hold any bound state or any narrow resonance, the single-particle representation of the first channel will be formed from real or complex energy scattering states exclusively.

The second channel mean-field differs from $V_2(r)$ used in Ref. \cite{1966Facio} by the presence of the Coulomb interaction $V_{coul}(r)$ 
 between the deuteron and triton and the central term having a surface Woods-Saxon (WS) form $V_{sws}(r)$,
\begin{eqnarray}
    V_2(r) &=&  \frac{-V_2}{1 
               + \left( \frac{R_2}{a_2}-1 \right)  e^{\frac{r-R_2}{a_2}}}  \\
    V_{sws}(r) &=&  \frac{- 4V_{sws}\, e^{\frac{r-R_2}{a_2}}}
      {\left(  1 + e^{\frac{r-R_2}{a_2}} \right)^2} \\
    V_{coul}(r) &=& \frac{e^2}{r}
\end{eqnarray}
with $V_2=52$ MeV, $a_2=0.85$ fm, $R_2=1.25$ fm and $e^2=1.43996508$ MeVfm. The strength of the surface WS was adjusted in order to have a resonance in the partial wave $s_{1/2}$. 
In our calculation we have taken  $V_{sws}=25$ MeV for which, the resonant energy is located $\varepsilon(s_{1/2})=0.125-i\, 0.101$ MeV ($k=0.0909-i\, 0.0319$ fm$^{-1}$). Then, the second channel model space is formed either from real energy scattering states or from a resonance and the appropriate complex contour.

 The coupling interaction  $V_{12}$ between the two channels is taken as in Ref. \cite{1966Facio},
\begin{eqnarray}
  V_{12}(r) &=& \frac{-V_{12}}{1 +
            \left( \frac{r}{a_{12}}-1\right) e^{\frac{r-R_{12}}{a_{12}}}}  
\end{eqnarray}
with $a_{12}=0.85$ fm, $R_{12}=1.0$ fm, and the coupling strength $V_{12}$ is a free parameter.

To study the poles of the CC model of $^5{\rm He}$ we use complex contours in the $k$ (first channel) and $p$ (second channel) planes. Table  \ref{table.contour} gives the vertices and the number of mesh points for each segment of the complex contour. When real contour is used, the imaginary component of $k_i$ or $p_i$ is zero. The $L^+$ contour is the collection of the line segments between the vertex points $k_i$ and $k_{i+1}$. There are $N_i$ mesh points in the segment determined by  $k_i$ and $k_{i+1}$. The vertices were chosen in order to uncover the energy region delimited by a polygon with vertices (in MeV) (0,0), (0,-0.5),(20,-0.5),(20,0) of the first channel and an identical shape shifted by $\Delta_2$ in the second channel. 

\begin{table}[ht] 
\begin{ruledtabular}
 \caption{\label{table.contour} Vertices ($k_i$ and $p_i$) and number of mesh points $N_i$ for each segment in the wave number complex plane. For further explanation see the text.}
\begin{tabular}{l|ll|ll}
 i&  \hspace{4mm} $k_i$ (fm$^{-1}$)    &$N_i$\hspace{4mm}  & $p_i$ (fm$^{-1}$) &$N_i$ \hspace{4mm} \\
\hline
 1& (0,0)               & 100 & (0,0)      & 100           \\
2&  (0.09816,-0.09816)   & 50 & (0.1201,-0.1201)  & 50 \\
3&  (0.8781,-0.01097)  & 40  & (1.0739,-0.01342) & 40  \\
4&  (1.2416,0)          & 40 & (1.5186,0)   & 40        \\
5&  (1.9631,0)          & 40 & (2.4011,0)  & 40         \\
 6& (8,0)               &  &(8,0)&
\end{tabular}
\end{ruledtabular}
\end{table}

First, we solved the coupled equations in a model space formed by the real energy scattering states in the first channel and use a complex contour in the second channel plus the $s_{1/2}$ resonance state. In this way poles in the Riemann sheets $(+,+)$ or $(+,-)$ can be obtained. As the $V_{12}$ strength is increased from zero, we found that a pole starting from $E=\varepsilon(s_{1/2})+\Delta_2$, moves upwards. We were not able to follow this pole beyond $V_{12} \simeq 13$ MeV. 
When  we use a complex contour for the first channel model space and the real energy contour for the second channel (poles of the Riemann sheets $(+,+)$ or $(-,+)$ can be obtained) then for small $V_{12}$ values 
we do not find any poles until we reach $V_{12} \simeq 14$ MeV. Then a  pole appears very close to the  real axis, and it moves downwards as the interaction increases. 

\begin{figure} 
  \includegraphics[width=0.3\textwidth,angle=-90]{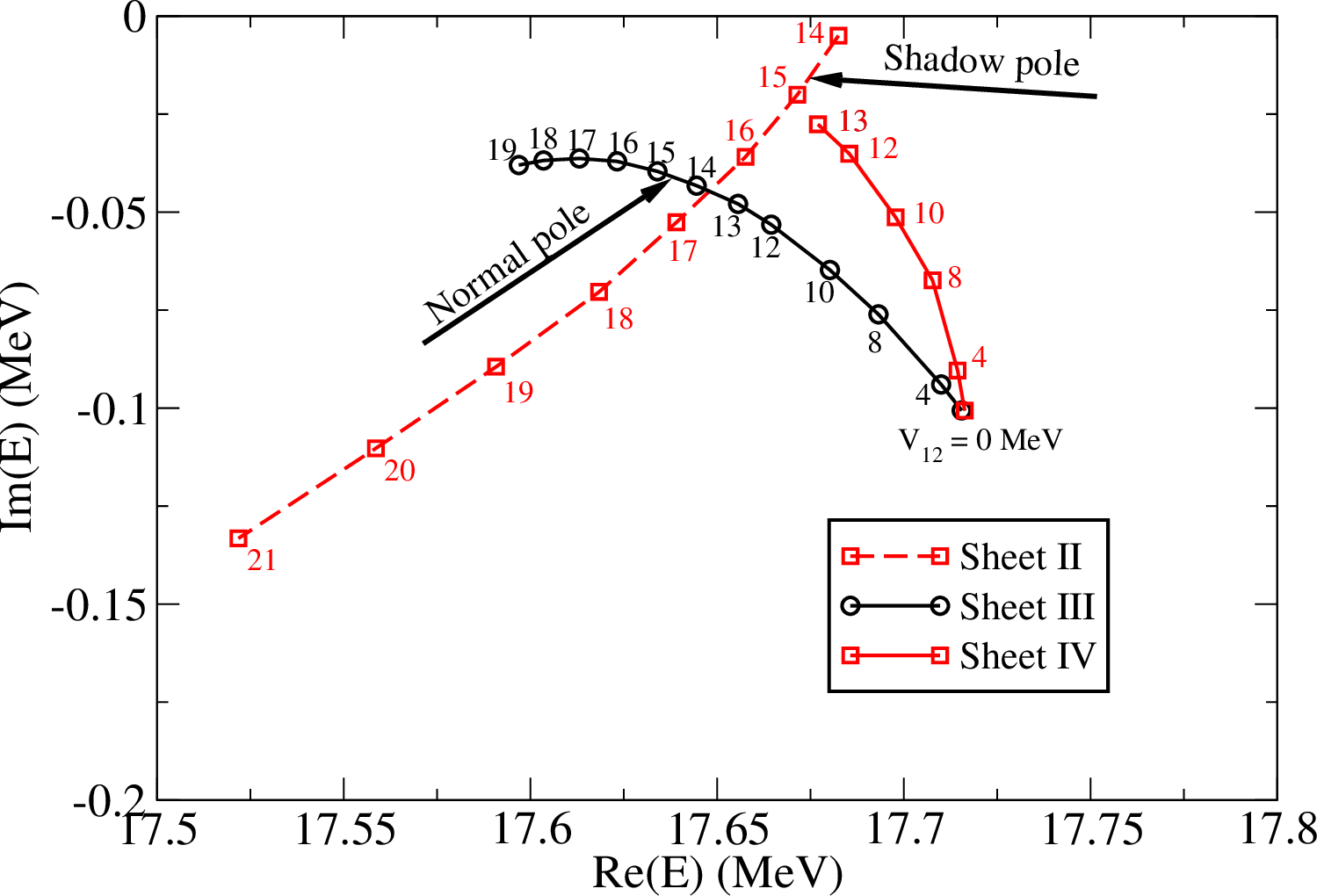}
  \caption[T]{\label{fig.move}
  (Color online) Trajectories of the physical and shadow poles as  functions of the interaction strength $V_{12}$. Circles (squares) denotes  the physical (shadow) poles.  The numbers next to the symbols indicate the values of  $V_{12}$ in MeV. Arrows point to the positions of the physical and shadow poles at the experimental strength $V_{12}=14.7$ MeV (see details in the text)}
\end{figure}

In Fig. \ref{fig.move} the pole  trajectories are shown as functions of the 
parameter $V_{12}$. Open squares connected by a full line are the results we got with the first model space and the poles are connected by dashed lines, for  second model spaces. In the the first model space the energy region of the fourth Riemann sheet is uncovered, while the second model space  the energy region of the second Riemann sheet is uncovered.

According to Ref. \cite{ede64}, if there is a resonant pole in any of the channels with no coupling, in the CC problem this pole will generate two poles in different Riemann sheets. In our model this implies that, when the channel interaction $V_{12}$ is gradually turned on,  one pole will appear in the fourth sheet (Im(k)$>$0, Im(p)$<$0) and the second pole in the third sheet (Im(k)$<$0, Im(p)$<$0). We have already found and discussed the movement of this latter one. 
When the complex contours are used in both channels (note that also the resonance must be included), we got a different resonance pole. This pole starts from the same position $\varepsilon(s_{1/2})+\Delta_2$ as the shadow pole for  $V_{12}=0$ and moves continuously toward the threshold $\Delta_2=17.59$ MeV.
Since this model space is characterized by Im(k)$<$0, Im(p)$<$0 it uncovers the fourth Riemann sheet, this pole corresponds to the physical resonance which appears in the sheet $U_{nd}$ of Ref. \cite{hal87-a}. The movement of this pole is also depicted in  \ref{fig.move}.

The behaviour of the pole trajectories displayed in Fig. \ref{fig.move}
clearly shows that, as the coupling strength increases, the shadow pole moves from the fourth Riemann sheet to the second one. The same type of pole migration was observed in \cite{csot93a} where a microscopic cluster model was used for the description of the  $J^\pi=3/2^+$ state.
From the pole trajectories we can notice that the real part of the energy of the shadow pole is always larger than that of the normal resonance pole if $0<V_{12}<19 {\rm MeV}$. This finding is in agreement with the result of the work \cite{csot93a}. 
 
The experimental energy of the normal pole is found to be at $0.048$ MeV \cite{tunl} with respect to the $^3$H-$d$ threshold and its width is $0.0745$ MeV. In our simple model
calculation we found that this pole position occurs in the third sheet for $V_{12}=14.7$ MeV.  At this strength the normal pole in our model is found at $0.048 - i\, 0.041$ MeV, i.e. $\Gamma=0.082$ MeV in good agreement with the experimental value. 
 At the same strength, the shadow pole is sitting in the second Riemann sheet with the following resonant parameters: $\varepsilon=0.088$ MeV and  $\Gamma=0.042$ MeV. The position of the shadow pole is close to that in other model calculations  ($\varepsilon\approx 0.082$ MeV, $\Gamma\approx 0.007$ MeV) \cite{csot93,hal87-a} the width however is overestimated.

\section{Conclusions} \label{sec:conclusions}
We have considered the exactly solvable CC problem of the Cox potential
and showed that by using the  Berggren 
expansion method we are able to reproduce all poles of the S-matrix even the shadow poles in agreement 
with the exact calculation. 
The proper choice of the complex contours is very 
important since it determines which CC states can be calculated using Berggren's basis. With suitably chosen contours we were able to calculate 
 poles on different Riemann sheets simultaneously.
  We also gave a numerical example in which
an anti-bound state was calculated using Berggren's basis 
formed only from bound and complex energy scattering states. 
The deviations of the numerical CC results from the exact results are within the accuracies of the
calculations of the single-channel basis states. 

We have shown a nuclear physics example where a shadow pole plays an important role. In the  $t+d\to\alpha+n$ fusion reaction 
the $\frac{3}{2}^+$ resonance state of the nucleus $^5{\rm He}$ is very important. We studied this state in a phenomenological two-channel model. In this example the emergence of two poles was observed, which are originating from a resonance in the $d+t$ channel. It was found that a shadow pole migrates between Riemann sheets if the coupling strength is varied and that at the physical strength, the normal and shadow pole parameters agree with previous findings.

\begin{acknowledgments}
Authors are grateful to Profs. B. Gyarmati, R. G. Lovas and A. Cs\'ot\'o for valuable discussions.
This work was supported  by the National Council of Research PIP-625 (CONICET, Argentina) and
by the Hungarian Scientific Research Fund-OTKA K112962.
\end{acknowledgments}

%


\end{document}